# HELEN: TRAVELING WAVE SRF LINEAR COLLIDER HIGGS FACTORY*

S. Belomestnykh[†,1], P. C. Bhat, A. Grassellino, S. Kazakov, H. Padamsee[2], S. Posen,
A. Romanenko, V. Shiltsev, A. Valishev, V. Yakovlev
Fermi National Accelerator Laboratory, Batavia, IL USA
[1]also at Stony Brook University, Stony Brook, NY USA
[2]also at Cornell University, Ithaca, NY USA

*Abstract*

Traveling wave SRF accelerating structures offer several advantages over the traditional standing wave structures: substantially lower $H_{pk}/E_{acc}$ and lower $E_{pk}/E_{acc}$, ratios of peak magnetic field and peak electric field to the accelerating gradient, respectively, together with substantially higher $R/Q$. In this paper we discuss how a linear collider Higgs Factory HELEN can be built using TW-based SRF linacs. We cover a plan to address technological challenges and describe ways to upgrade the collider luminosity and energy.

## INTRODUCTION

Recently completed 2021 U.S. Community Study on the Future of Particle Physics (Snowmass 2021) identified the most important scientific questions in High Energy Physics for the following decade [1]. One of the goals emphasized by the Energy, Accelerator, and Theory Frontiers is "to position the U.S. HEP program to support construction of a Higgs Factory as early as 2030."

A standing wave linac TESLA superconducting radio frequency (SRF) technology has been developed for the International Linear Collider (ILC). This technology is well-established and mature ("shovel ready") as demonstrated by recent large-scale projects European XFEL in Hamburg, Germany, and LCLS-II/LCLS-II-HE at SLAC in the USA. At present, multi-cell TESLA cavities are limited to accelerating gradients of ~ 40 – 45 MV/m. With an advanced cavity treatment and more efficient cell geometries, the community hopes to advance the limit to ~ 50 – 55 MV/m [2, 3]. Further improvements require alternative approaches, e.g., different acceleration schemes or superconductors with higher RF critical magnetic field.

In this paper we discuss advantages of using a traveling wave (TW) SRF acceleration scheme and how such a scheme can be applied to a recently proposed Higgs-Energy LEptoN (HELEN) collider [4, 5]. We present an R&D plan and ways to upgrade the collider.

## TRAVELING WAVE SRF TECHNOLOGY

Operation of superconducting RF structures in traveling wave regime was first studied in mid-1960s with a detail analysis published by Neal in 1968 [6]. As the SRF structure attenuation is extremely low, one must avoid wasting electromagnetic power at the end of the structure. This is done by implementing a resonant ring set up, in which the residual RF power at the end of the accelerating cavity is fed back to the input end via a waveguide, where it is combined with the source power and fed back to the accelerator.

### Structure Layout, Advantages and Drawbacks

RF power source can be connected to the feedback waveguide in several different ways: via a directional coupler, or either one, two or more RF power couplers. The one- and two-coupler schemes were examined in detail in [7]. Fig.1 illustrates the two-coupler scheme. A 15-cell cavity operating in a $\pi/2$ mode is coupled to a rectangular waveguide at both ends, thus creating a resonant ring. An adjustable matcher with reflection coefficient Γ is used to compensate reflection from the TW section. The scattering matrix formalism was used for the system analysis. For this purpose, the structure is sub-divided into several sections, each characterized by its own scattering matrix. The sections' boundaries are indicated with dashed lines in the figure, vectors $a_n$ and $b_n$ represent the incident and reflected waves.

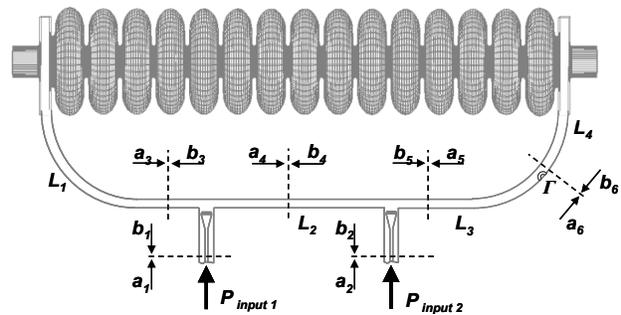

Figure 1: Two-coupler model of the resonant travelling wave feedback ring and a approximately one-meter long TW cavity.

The TW structure provides the following benefits with respect to the conventional standing wave SRF cavity [8]:

1. Higher transit time factor ($T \sim 1$) and higher acceleration gradient for the same peak surface RF magnetic field. For an ideal structure with small aperture $T(\varphi) \sim \sin(\varphi/2)/(\varphi/2)$, where $\varphi$ is the RF phase advance per cell. Then the acceleration

---
* This work was produced by Fermi Research Alliance, LLC under Contract No. DE-AC02-07CH11359 with the U.S. Department of Energy. Publisher acknowledges the U.S. Government license to provide public access under the DOE Public Access Plan DOE Public Access Plan.
† sbellomes@fnal.gov

gradient increase compared to the standing wave structure operating in the $\pi$ mode is

$$Gain = T(\varphi)/T(\pi) = (\pi/\phi) \times \sin(\phi/2) \quad (1)$$

The gain in the accelerating gradient of the ideal $\pi/2$ traveling wave accelerating structure is 41.4%.

2. High stability of the field distribution along the structure with respect to geometrical perturbations. This allows a much longer structure length, which would be limited by manufacturing and cavity treatment technology rather than physics limited and would result in (i) potentially reduced number of input couplers, limited by the coupler capability, and (ii) fewer gaps between accelerating structures, hence better real estate gradient.
3. The TW structure has no trapped modes in the lower dipole mode passband. Only two HOM dampers per long TW structure are required.
4. A transverse kick caused by the fundamental RF power and HOM couplers is not an issue for long TW structure, because the number of the couplers is small. In addition, the couplers may be optimized for minimizing the transverse kick.

There are some drawbacks though. First, the structure includes a high-power feedback waveguide and lager number of cells (2 times for $\pi/2$ TW), which will make the structure fabrication more complicated. Second, the feedback waveguide requires careful tuning to compensate reflections along the resonant ring and thus obtain a pure traveling wave regime at the desired frequency. Third, high circulating power in the resonant ring must be demonstrated. Despite these drawbacks, the advantages of the traveling wave approach are significant to warrant further development.

A traveling wave SRF cavity optimization study [9] demonstrated that for an aperture radius $R_a = 25$ mm (smaller that 35 mm of the TESLA cavity geometry) and phase advance of 90°, one can achieve $H_{pk}/E_{acc} = 28.8$ Oe/(MV/m) with $E_{pk}/E_{acc} = 1.73$. Since $H_{pk}/E_{acc}$ is 42.6 Oe/(MV/m) for the TESLA structure, the TW structure has reduced the critical parameter $H_{pk}/E_{acc}$ by almost a factor of 1.5. At the same time, the peak electric field ratio is smaller than TESLA cavity value of 2.0 and we gain a factor of 2.1 in $R/Q$ although losing in the geometry factor by 1.45 (186 Ohm vs. 270 Ohm). Thus, for the same accelerating gradient the TW cavity is more cryogenically efficient than TESLA cavity. A high group velocity in the TW mode increases the cell-to-cell coupling from 1.8% for the TESLA structure to 2.3%. This confirms that TW structures have less sensitivity to cavity detuning errors, making the tuning easier, despite having more cells.

## HELEN: A HIGGS FACTORY BASED ON TW SRF

Padamsee [10] considered several options for the ILC energy upgrade. One of the options is to use the traveling wave SRF operating at ~ 70 MV/m. However, providing that there is time for the technology development prior to beginning of the project, we proposed to use this advanced SRF option from the very beginning for a compact $e^+e^-$ linear collider HELEN baseline [4]. If sufficient funding for R&D is available, the traveling wave SRF could be demonstrated in a fully developed prototype cryomodule within approximately 5 years, after which the project could be initiated. Other than TW SRF structures, HELEN is very similar to ILC, as one can see in the collider layout (Fig. 2) and the parameter list presented in Table 1. However, due to higher accelerating gradient (and larger fill factor of 80.4% vs. 71% for ILC), HELEN offers ~ 37% cost saving for the main linac. All proposed ILC luminosity upgrade scenarios (see, e.g., [7]) are applicable to HELEN.

Table 1: HELEN parameters

| Parameter | Value |
|---|---|
| Center of mass energy | 250 GeV |
| Collider length | 7.5 km |
| Peak luminosity | 1.35×10$^{34}$ cm$^{-2}$s$^{-1}$ |
| Repetition rate | 5 Hz |
| Bunch spacing | 554 ns |
| Particles per bunch | 2×10$^{10}$ |
| Bunches per pulse | 1312 |
| Pulse duration | 727 µs |
| Pulse beam current | 5.8 mA |
| Bunch length, rms | 0.3 mm |
| Crossing angle | 14 mrad |
| Crossing scheme | crab crossing |
| RF frequency | 1300 MHz |
| Accelerating gradient | 70 MV/m |
| Real estate gradient | 55.6 MV/m |
| Total site power | 110 MW |

We have identified locations for a possible future linear collider at Fermilab: two 7-km diagonal options and a 12-km footprint with North-South orientation extending outside the site boundary but with the Interaction Region (IR) on site, see details in [11, 5]. A 250-GeV HELEN Higgs factory (HELEN-250) could potentially fit along either of two diagonals after further optimization of the collider. The North-South footprint can accommodate not only HELEN-250 but would also allow extension of the main linacs to the center-of-mass energy of 500 GeV.

## TRAVELING WAVE SRF R&D

We anticipate that up to 2-meter-long traveling wave accelerating structures could be feasible, resulting in fewer cavity-to cavity transitions in the linac and hence to larger real-estate gradients. However, the R&D program focuses on a one-meter structure due to limitations of existing facilities.

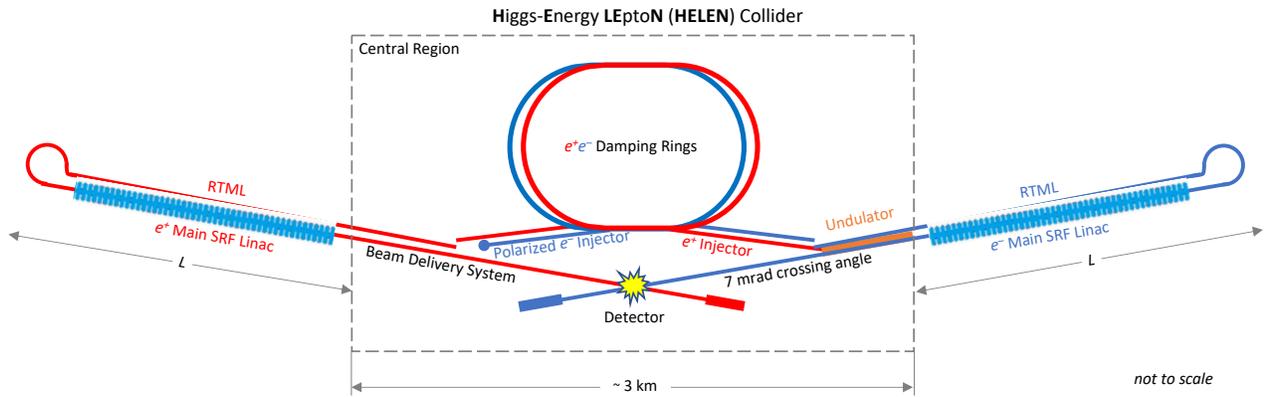

Figure 2: Conceptual layout of the HELEN collider.

A proof-of-principle technology demonstration effort started several years ago with fabricating and testing a single-cell cavity with a feedback waveguide [12]. The test proved that attaching a waveguide to the cell does not degrade the cavity performance. The next step was to fabricate and test a 3-cell cavity with the waveguide. The cavity (Fig. 3) was designed and fabricated [13]. Following fabrication, it received BCP treatment and is currently undergoing tuning at Fermilab. The goal for this step is to prove that the cavity can be tuned to support the traveling wave and demonstrate traveling wave performance in a vertical test.

Upon successful testing of the 3-cell cavity, we plan to proceed with further demonstrations of feasibility of the traveling wave technology:

- Design, build and test a proof-of-principle multi-cell, half-meter- to one-meter-long 1.3 GHz TW cavity and demonstrate accelerating gradient of ~70 MV/m.
- Adapt an advanced cavity treatment technique so that high Q ~ $10^{10}$ can be achieved at high gradient.
- Design, build and test several dressed prototype cavities, demonstrate performance required for the HELEN collider.
- Design and build a prototype cryomodule.
- Verify the cryomodule performance without beam on a test stand and with beam at the Fermilab's FAST facility.

Beyond the demonstration of TW technology, the collider R&D program (if funded) will pursue the following tasks:

- Design and optimization of the HELEN linear collider accelerator complex.
- Confirm the physics reach and detector performance for the HELEN beam parameters.
- Publish Conceptual Design Report as modification of the ILC design in 2–3 years.
- Prepare Technical Design Report after demonstrating the cryomodule performance, in ~5 years.

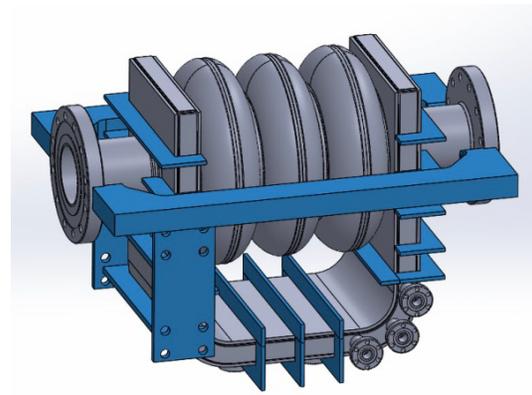

Figure 3: 3D model of 3-cell traveling wave cavity [13].

## SUMMARY

In this paper we described a Higgs Factory linear collider HELEN based on a traveling wave SRF technology. While this technology Is not fully demonstrated yet, the expertise and technological advances accumulated by the world SRF community – in particular, at the U.S. laboratories and universities (Cornell, Fermilab, JLAB, SLAC, ...) – would allow rapid development, prototyping, and testing of SRF cavities and cryomodules based on the traveling wave SRF. After the full potential of the technology is demonstrated, it can find other applications as well, e.g., 8-GeV SRF linac option for the ACE project at Fermilab and future SRF-linac based compact FELs. Fermilab has capabilities that support the full cycle of R&D, production, and verification (including testing cryomodules with beam) at the SRF accelerator test facilities and FAST linac.

If given high priority, the construction of the HELEN-250 collider could start as early as 2031–2032 with first physics in ~2040. The HELEN collider can be upgraded to higher luminosities in the same way as was proposed for the ILC or to higher energies (up to 500 GeV at Fermilab site) by extending the linacs.